\documentclass[12pt, journal]{IEEEtran}
\usepackage[T1]{fontenc}
\usepackage{textcomp}
\usepackage[margin=1in]{geometry}
\usepackage{graphicx}
\usepackage{caption}
\usepackage{subcaption}
\usepackage{setspace}
\usepackage{parskip}
\usepackage{fancyhdr}
\usepackage{amsmath}
\usepackage{mathtools}
\usepackage{todonotes} 
\usepackage{slashbox}

\graphicspath{{figures/}}

\title{Predicting passenger loading level on a train car: A Bayesian approach}
\author{ARCON Corporation}
\date{November 30, 2017}

\begin{document}
\onecolumn
\newcommand{\massbogie}{m_b}
\newcommand{\masscar}{m_c}
\newcommand{\kbogie}{k_b}
\newcommand{\kcar}{k_c}
\newcommand{\cbogie}{c_b}
\newcommand{\ccar}{c_c}
\newcommand{\tf}{\text{H(s)}}
\title{Predicting Passenger Loading on Train Cars using APC data: A Bayesian approach}
%
%
%

\author{Peter~Khomchuk,~\IEEEmembership{}
        Saurav~R.~Tuladhar,~\IEEEmembership{}
        and~Siva~Sivananthan,\\~\IEEEmembership{ARCON, Waltham, MA}
\thanks{Work performed under Phase I SBIR grant from the Volpe National Transportation Systems Center, US Department of Transportation}}

%
%

\maketitle



\pagestyle{plain}
\pagenumbering{arabic}
\setcounter{page}{1}
\section{Background}
\label{ch:background}
Crowding in train cars is increasingly a major concern for transit agencies. From the perspective of the passengers and the transit agencies, overcrowding of the train cars has several negative consequences such as: (i) extended duration of passengers boarding and alighting which leads to longer dwell times, (ii) subsequent disruption of the headway and the schedule, and (iii) passenger dissatisfaction (e.g. increased stress and lack of privacy). Moreover, overcrowding during peak service hours also indicates inadequate infrastructure to meet the passenger demands. Realizing the importance of the crowding issue, transit agencies have developed measures to assess the crowding levels \cite{wmata2014, zheng2013}. The Transit Capacity and Quality of Service Manual provides guidelines on thresholds for crowding in transit systems in the United States \cite{tcqsm2003}.

\subsection{Passenger loading in train cars}
Ideally the passengers should be uniformly distributed among the cars in a train consist. However, passenger distribution has been observed to be significantly uneven among the train cars during peak hours. Train cars that are closer to the entrances or major escalators and stairway landings are  heavily loaded compared to the cars further away. The uneven loading of train cars can lead to overcrowding of certain cars while other cars remain only sparsely occupied if not empty. It is in the interest of the transit agencies to maintain balanced passenger distribution in the train cars to maximize system efficiency and passenger comfort. One approach to mitigate the uneven passenger distribution is to provide prior information about the train car crowding levels to passengers waiting at the next station. The waiting passengers can then choose to position themselves to board the car that is expected to be less crowded, leading to relatively even passenger distribution through the train consist. 


\subsection{Measuring passenger crowding}
In the past, most transit systems relied on manual counting for collecting data and monitoring crowding conditions in the train cars. However, recent advancements in automatic passenger counter (APC) technology resulted in a variety of devices that automatically measure the number of passengers entering or exiting a train car \cite{reuter2003passenger}. Such APC devices include infrared or CCTV based counters, treadle mats, and vehicle weighing devices. 

The CCTV video or infra-red (IR) image based APCs are known to provide  the most robust and accurate passenger counts. These vision-based APCs can be up to 97\% accurate and require minimal changes after installation. The CCTV or IR sensors are installed at each door in the train car. They provide directional information, such that passengers entering and leaving a car can be counted separately. The vision-based methods are robust to light and brightness conditions, differences in passengers height and appearance. Since these sensors count only the number of boarding and alighting passengers, and not the total number of passengers in the car at a given time, the train crowding information estimated using these sensors is susceptible to an accumulating error which increases over time, especially after visiting stations with a high passenger flow.

Another set of APC technology is based on directly measuring changes in the weight of a train car due to boarding and alighting passengers. Most of the modern train cars are equipped with electronic weighing sensors which provide information to the train braking system. The data from these sensors can be used to infer passenger load count. Experiments in the Copenhagen rail network show that the weight-based approach provides passenger counts comparable to the infra red-based APCs \cite{Nielsen}. Another approach to weight-based APC uses weight measuring transducers installed on rail lines between two stations. As the train cars pass over the weight sensors, the weights of train cars can be measured. 

The proliferation of WiFi and bluetooth capable devices among the public makes them a suitable candidate for passenger counting. Researchers are developing reliable methods to infer the number of passengers inside a train car using the number of bluetooth devices detected \cite{Maekawa}. However, WiFi and bluetooth based APC technology is not robust enough for systematic deployment for large transit systems \cite{wmata2014}.

In general, the current set of APC technologies are well suited for use in buses to count passengers. Using these APCs on trains may be more difficult due to complications unique to railcar entry. However, some agencies, including Berlin and Hamburg, are using APCs successfully in their trains. Berlin and Hamburg use doorway infrared devices to count passengers in the train cars. 

The data from the APCs are currently being used by transit agencies for various applications such as on-time-performance, revenue analysis, updating schedules and upgrading capacity. But to our knowledge, no transit agency in the US is using the count data to provide crowding information to the waiting passengers.

\section{Crowding Information Collection and Dissemination (CICD) System}
\label{sec:cicd}
Crowding information collection and dissemination (CICD) system is a solution for providing information to awaiting passengers at a station about availability of space in each car of the arriving train. The CICD system measures the passenger loading on each train car and predicts the expected crowding level after the on-board passengers alight at the next station. The CICD system provides the expected crowding level information to the the passengers waiting at the arriving station. The CICD system is designed to have three modules as shown in Figure~\ref{fig:cicd_system}. The individual modules are described below:

\begin{figure}[!ht]
\centering
\includegraphics[width=\textwidth]{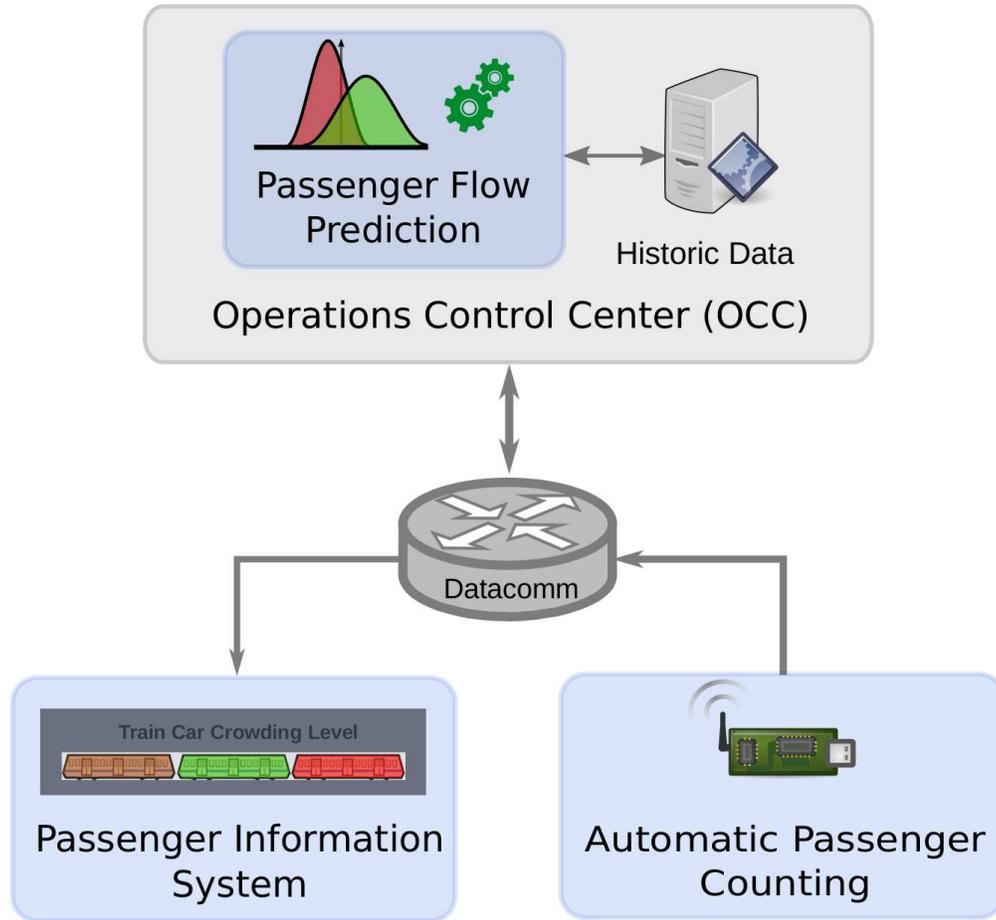}
\caption{Proposed CICD system concept}
\label{fig:cicd_system}
\end{figure}

\begin{itemize}
\item {\it Automatic Passenger Counting (APC) module}: The APC module provides an estimated on-board passenger count on each train car. The APC module can either use the installed APC devices or an accelerometer based APC as proposed below. The accelerometer based APC aims to be an economical passenger counting approach compared to existing computer vision and infra-red based technologies.

\item {\it Passenger Flow Prediction (PFP) module}: The PFP module uses Bayesian inference to predict the boarding and alighting flow of passengers in the train cars. The predictive algorithm combines historical passenger flow data and the current measurements from the APC module to predict the passenger flow. The predicted alighting passenger flow at the next station is used to generate the expected crowding level for each train car. The PFP module will be housed inside the transit agencies' Operations Control Center (OCC). From the OCC, the expected crowding level information is forwarded to the relevant stations. 

\item {\it Passenger Information System (PIS) module:} The PIS delivers information about the train car space availability to the passengers in a timely and comprehensible manner. The PIS module comprises of existing infrastructure used to communicate transit information to the passengers. Additionally, the crowding information will be made available to passengers via websites and smartphone apps. 
\end{itemize}
\section{Crowding level prediction}
\label{ch:pfp}
The primary goal of the CICD system is to provide advance information to passengers waiting at the station on the predicted crowding level on each car of the arriving train. To achieve this goal, an accurate crowding level prediction for each train car is essential. ARCON proposed to develop an algorithm to predict crowding level by combining the current passenger loading data, historical passenger data and the predicted passenger flow on each train car. This algorithm is a part of the Passenger Flow Prediction (PFP) module in the CICD system. As a part of the Phase I effort, ARCON has been working on developing a prototype PFP algorithm. The section below provides an overview of the algorithm design approach pursued during the current PoP.

\subsection{Statistical modeling} 
ARCON's approach to crowding level estimation for each car is based on statistical modeling of the passenger flow and is related to the known problem of origin-destination matrix (ODM) estimation in transit systems.  We assume that it is not possible to explicitly measure the amounts of boarding and alighting passengers, and the only available measurements are the readings from the car level APCs. This is a highly under-determined problem with a large number of unknowns and a much fewer measurements. Thus our approach will rely heavily on available historical data and a number of statistical models.

\subsubsection{Origin-destination matrix}
A route-level ODM indicates the number of passengers traveling from a specific origin station to a specific destination station. Such ODMs are widely used by transit agencies for route planning and scheduling. Usually, the origin-destination data is estimated from the station-level passenger counts obtained from automated fare collection (AFC) systems. 

The route-level ODMs are time dependent and provide only average estimates with a time resolution on a scale of half an hour to several hours. A typical ODM has a form shown in Table \ref{tb:odm}. For a route with $M$ stations the ODM is an $M \times M$ square matrix. An $(i,j)$ entry in the matrix contains the number of passengers traveling from the station $i$ to the station $j$. The route-level ODM considers traveling in one direction only and as such the ODM is strictly upper triangular. As a result, the sum across $i^{\rm th}$row equals the total number of passengers that boarded at the $i^{\rm th}$ station, and the sum across $j^{\rm th}$ column equals to the number of passengers alighting at the $j^{\rm th}$ station.

\begin{table}[!ht]
\centering
\caption{General form of an origin-destination matrix (ODM).}
\label{tb:odm}	
	\begin{tabular}{|c|c|c|c|c|c|}
    	\hline
         \backslashbox{Origin}{Destination} & station 1 & station 2 & station 3 &  $\ldots$ & station N \\
         \hline
         station 1 & & $b_{1,2}$ & $b_{1,3}$ & $\ldots$ & $b_{1,N}$\\
         \hline
         station 2 & & & $b_{2,3}$  & $\ldots$ & $b_{2,N}$ \\
         \hline
         station 3 & & & & $\ldots$ & $b_{3,N}$ \\
         \hline
         $\vdots$ & $\vdots$ & $\vdots$ & $\vdots$ & $\ddots$ & $\vdots$\\
         \hline
         station N & & & & $\ldots$ &\\
         \hline
    \end{tabular}	
\end{table}

Assuming that the passengers arrive to a station $i$ at a constant rate $\lambda_i$, the number of passengers, $b_i$, waiting to board at the $i^{\rm th}$ station can be modeled as a Poisson distributed random variable
\begin{equation}
	b_{i} = \sum_{j=1}^M b_{i,j} \sim Poiss( \lambda_i \Delta t_i )
\end{equation}
where $b_{i,j}$ is a number of passengers traveling from station $i$ to $j$, and $\Delta t_i$ is a time between two consecutive trains at the $i^{\rm th}$ station. The proportions (probabilities) of passengers going from the $i^{\rm th}$ station to all subsequent $M_i = M - i$ stations  $p_{i,j} = b_{i,j}/b_i$, $j = i+1,\ldots,M$, can be modeled as a Multinomial probability distribution. This means that each $b_{i,j}$ is also Poisson distributed such that $b_{i,j} \sim Poiss( \lambda_i \Delta t_i p_{i,j} )$. Similarly, the number of alighting passengers at the $j^{\rm th}$ station is
\begin{equation}
	a_{j} = \sum_{i=1}^{j-1} b_{i,j}
\end{equation}
and the number of on-board passengers when the train reaches the $j^{\rm th}$ station is
\begin{equation}
	o_{j} = \sum_{j'=j}^{M} a_{j}
\end{equation}
Thus, the ODM describes a passenger flow model according to which the passengers arrive to the $i^{\rm th}$ station with a rate $\lambda_i$ and then choose their destination station with a probability $p_{i,j}$. The number of boarding, alighting, and on-board passengers are Poisson distributed, with the parameters that could be determined from the ODM.


\subsubsection{Model for car crowding level}
ARCON proposes to use available route-level ODMs to come up with ODMs for each car of a particular train. Such car-level ODMs will be updated in real-time with APC data. The car-level crowding information will be available from these car-level ODMs as an estimate of the number of on-board passengers.

Initial estimates of the car-level ODMs will be obtained from the corresponding route-level ODM by taking into account two factors: (i) station layouts and (ii) the probability of a passenger being committed to choosing a boarding position according to the exit location at his/her destination station. 

Several studies have found that the passenger distribution on the platform depends on the entry and exit locations in the station platform. Among the total passengers entering the platform, a certain proportion is committed to choose a particular train car for boarding. Most of these committed passengers choose a specific train car to minimize the distance to the exits at their destination station \cite{kim2014passengers}. Hence at any given station, the committed passengers will most likely position themselves on the platform location corresponding to the train cars which minimizes the distance to exit locations at subsequent stations. The remaining non-committed passengers tend to distribute themselves along the platform such that there are more passengers near the entry location and fewer passengers further away. 

\begin{figure}[!ht]
\includegraphics[width=\textwidth]{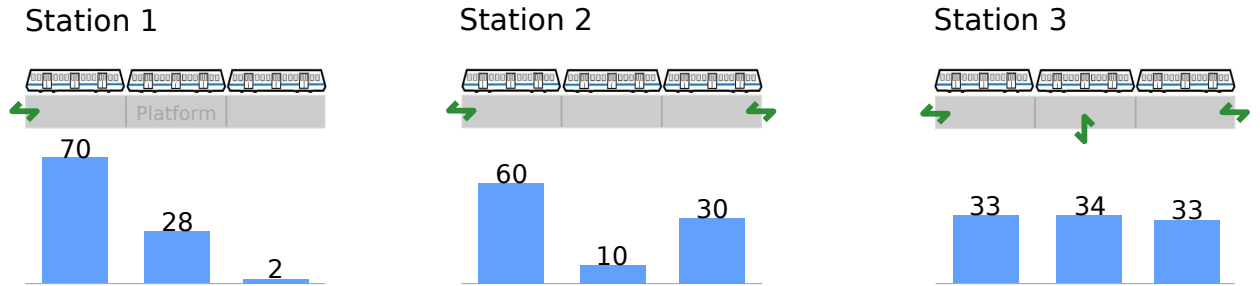}
\caption{Station layout model and non-committed passenger distribution along the platform. Points of access to the platform are marked in green.}
\label{fig:station-model}
\end{figure}

In our model we assume that the boarding passengers position themselves into one out of $K$ possible positions on the platform where $K$ is the number of cars in the train. In other words, prior to a train arrival, a passenger decides which car he/she is going to board. This decision is final and it is not changed. The number of passengers at the $i^{\rm th}$ station that are waiting to board the $k^{\rm th}$ car, and whose destination is the $j^{\rm th}$ station is
\begin{equation}
	b_{i,j}^k = b_{i,j}^{k,c} + b_{i,j}^{k,nc}
\end{equation}
where $b_{i,j}^{k,c}$ is a number of committed passengers who choose the $k^{\rm th}$  car because at their destination at the $j^{\rm th}$ station they want to be closer to an exit, and $b_{i,j}^{k,nc}$ is the number of not-committed passengers who choose the $k^{\rm th}$ car because of the layout of the $i^{\rm th}$ station. Here, the superscripts $c$ and $nc$ indicate respectively committed and not-committed passengers. Further we define a number of probabilities:
\begin{itemize}
\item The probability, $p_c$, of a passenger being a committed passenger. This probability can be determined by conducting a survey.
\item The probability, $p_i^{k,nc}$, of a not-committed passenger choosing the $k^{\rm th}$ boarding position at an $i^{\rm th}$ station. This probability depends on the $i^{\rm th}$  station's layout as shown in Fig \ref{fig:station-model}. A number of studies and surveys have shown that $p_i^{k,nc}$ decays proportionally to the distance from the boarding position $k$ to the platform's access points
\begin{equation}
\label{eq:boarding_dist}
p_i^{k,nc} = \frac{1}{Z_i} \sum_{ l \in \left\{ L_i \right\} } \lvert k - l\rvert^{-\xi_i}
\end{equation}
where $L_i$ is a set of access points (stairs, elevators, escalators, etc.) at the $i^{\rm th}$ station, $\xi_i > 0$ is a constant that determines the rate of decay, and $Z_i$ is a proportionality constant that guarantees that the probabilities over all boarding positions sum to one. Such distribution guarantees that most passenger wait for a train near the entrances to the platform, and that their number decreases as the boarding position gets further away from the entrances.

\item Finally, $p_j^{k,c}$ is the probability that a committed passenger who is traveling to the station $j$, chooses the $k^{\rm th}$  boarding position. This probability does not depend on the origin station and is determined only by the layout of the destination station. It can be uniformly divided between the positions that correspond to the exits at the $j^{\rm th}$  station, and be equal to zero for all other positions.
\end{itemize}

Using these probabilities the quantities $b_{i,j}^{k,c}$ and $b_{i,j}^{k,nc}$ will have the following distributions
\begin{equation}
	b_{i,j}^{k,nc} \sim Poiss( \lambda_i \Delta t_i p_{i,j} (1-p_c) p_{i}^{k,nc} )
\end{equation}
and
\begin{equation}
	b_{i,j}^{k,c} \sim Poiss( \lambda_i \Delta t_i p_{i,j} p_c p_{j}^{k,c} )
\end{equation}
Now the elements of the car-level ODM for the $k^{\rm th}$ car can be defined in terms of the above described probabilities, the passenger arrival rate, and the train headway time
\begin{equation}
\label{eq:prior}
	b_{i,j}^k \sim Poiss(\lambda_i \Delta t_i p_{i,j} ( p_{i}^{k,nc}(1-p_c) + p_{j}^{k,c} p_c) )
\end{equation}
Notice, since we assume that the passengers do not change their boarding positions after arriving to the station, each one out of $K$ car-level ODMs can be considered independently from one another.

The number of on-board passengers in the $k^{\rm th}$ car when the train arrives to the $j^{\rm th}$  station is a sum of Poisson distributed random variables
\begin{equation}
\label{eq:on_board}
o_j^k = \sum_{i = 1}^{j-1} \sum_{j'=j}^M b_{i,j'}^k
\end{equation}
and thus is also Poisson. A rationale behind this equation is that the number of passengers in a car $k$ is a sum of all passengers who boarded this car at any station preceding station $j$, and whose destination is $j$ or any stations following station $j$. An iterative procedure will update the distribution of $o_j^k$ using real-time APC data from the $k^{\rm th}$ car.

Further, we define the car crowding level for the $k^{\rm th}$ car and the $i^{\rm th}$ station as the number of passengers that remain in the car after the train has stopped and the alighting passengers got off
\begin{equation}
\label{eq:crowding}
\omega_j^k = o_j^k - a_j^k = o_j^k - \sum_{i=1}^{j-1} b_{i,j}^k = \sum_{i = 1}^{j-1} \sum_{j'=j + 1}^M b_{i,j'}^k
\end{equation}
where
\begin{equation}
\label{eq:alighted}
a_j^k = \sum_{i=1}^{j-1} b_{i,j}^k
\end{equation}
is the number of passenger alighted from the $k^{\rm th}$ car at the $j^{\rm th}$ station.

\subsection{Bayesian estimation approach}

As can be observed from equations (\ref{eq:on_board}) and (\ref{eq:crowding}), the entries $b_{i,j}^k$ of the car-level ODMs fully determine our quantities of interest i.e. the number of passengers on board and the crowding levels. Thus, ARCON proposes the following estimation scheme: given available historical data and the statistical model in (\ref{eq:prior}), come up with a prior distribution for the elements of the car-level ODMs, then at each station using APC measurements and the model in (\ref{eq:on_board}) update the ODM entries. Finally, using (\ref{eq:crowding}) calculate the crowding levels. Such an iterative procedure that follows the prior-measurement-update steps falls under Bayesian paradigm and can be effectively solved using Bayesian estimation techniques. 

Since we assume that after arriving to a station the passengers do not change their chosen boarding positions, in the following discussion we consider the ODM estimation for a $k^{\rm th}$ car. The same procedure applies to all other cars.

Using (\ref{eq:on_board}) the following matrix equation can be setup
\begin{equation}
{\bf{ o } }_j^k = {\bf{ A } }_j{\bf{ b } }_j^k
\label{eq:onb-matrix}
\end{equation}
where ${\bf{o}}_j^k = [ o_2^k,\ldots, o_j^k]$ is a $(j - 1) \times 1$ vector of the numbers of on-board passengers starting from the first station all the way to the current station $j$, ${\bf{ b }}_j^k = [b_{1,2}^k,\ldots,b_{1,M}^k, b_{2,3}^k,\ldots,b_{2,M}^k,\ldots]$ is a $\frac{1}{2}M(M-1)\times 1$ vector of entries above the main diagonal of ODM, and ${\bf{ A }}_j$ is a matrix of ones and zeros. For example, for a line with 3 stations Eq.~\ref{eq:onb-matrix} has the following form
\begin{equation}
\begin{bmatrix}
o_2^k\\
o_3^k
\end{bmatrix}
=
\begin{bmatrix}
1 & 1 & 0\\
0 & 1 & 1
\end{bmatrix}
\begin{bmatrix}
b_{1,2}^k\\
b_{1,3}^k\\
b_{2,3}^k
\end{bmatrix}
\end{equation}
Notice that in this example and in the definition of vector ${\bf{o}}_j^k$ in (\ref{eq:onb-matrix}) the element $o_1^k$ is missing. This is because $o_j^k$ is the number of passengers in the $k^{\rm th}$ car prior to arriving at the station $j$. Since the train is empty before the first station, $o_1^k = 0$. For convenience this element is omitted from ${\bf{o}}_j^k$.

In practice, the APC measurements of the on-board passenger count usually have errors. The errors can be modeled as random noise added to the true on-board counts. The APC measurements at the $k^{\rm th}$ car can be modeled as follows
\begin{equation}
\label{eq:linear_form}
{\bf{ o } }_j^k = {\bf{ A } }_j{\bf{ b } }_j^k + {\bf{ w }}_j^k
\end{equation}
where ${\bf{ w }}_j^k$ is an additive random noise. The choice of the APC noise distribution and its parameters such as mean and variance will define the quality of the APC measurements. For example, the camera based APCs might have low bias and variance since they were shown to be very accurate, while the proposed accelerometer-based APC approach might have a higher bias and variance.

Adopting Bayesian inference framework and using the distribution in (\ref{eq:prior}) as the prior for the elements of vector ${\bf{ b } }_j^k$ in (\ref{eq:linear_form}), we can estimate the posterior distribution of the entries of the car-level ODM matrix given new APC measurements as
\begin{equation}
\label{eq:posterior}
p( {\bf{ b } }_j^k | {\bf{ o } }_j^k) \sim p( {\bf{ o } }_j^k | {\bf{ b } }_j^k ) p({\bf{ b } }_j^k ) = p( {\bf{ o } }_j^k | {\bf{ b } }_j^k ) \prod_{i=1}^{M-1} \prod_{j'=i+1}^M p(b_{i,j'}^k)
\end{equation}
Here $p({\bf{ b } }_j^k )$ is a prior distribution and $p( {\bf{ o } }_j^k | {\bf{ b } }_j^k )$ is the likelihood function defined by the chosen APC noise model. The prior $p({\bf{ b } }_j^k )$  is equal to the product of Poisson distributions in (\ref{eq:prior}) for individual entries of the ODM matrix since the ODM entries are statistically independent.

The actual values of the vector ${\bf{ b } }_j^k$ can be estimated from the posterior distribution using maximum \textit{a posteriori} (MAP) estimator

\begin{equation}
\label{eq:map1}
\hat{\bf{ b } }_j^k = \max_{{\bf{ b }}_j^k}p( {\bf{ b } }_j^k | {\bf{ o } }_j^k) =  \max_{{\bf{ b }}_j^k}p( {\bf{ o } }_j^k | {\bf{ b } }_j^k ) p({\bf{ b } }_j^k )
\end{equation}

If the APC noise is modeled as a Gaussian distribution with zero mean and a diagonal covariance $\sigma^2 {\bf{I}}$, where $\sigma$ is a standard deviation of the APC noise, the MAP estimate of ${\bf{ b } }_j^k$ is a solution to the following optimization problem
\begin{equation}
\label{eq:map}
\hat{\bf{ b } }_j^k = \max_{{\bf{ b }}_j^k}\mathcal{N}\left({\bf{ A } }_j{\bf{ b } }_j^k,\sigma^2 {\bf{I}} \right) \prod_{i=1}^{M-1} \prod_{j'=i+1}^M p(b_{i,j'}^k)
\end{equation}
where $\mathcal{N}\left({\bf{\mu}},{\bf{\Sigma}}\right)$ is a multivariate Gaussian distribution with mean vector ${\bf{\mu}}$ and covariance $\bf{\Sigma}$. This optimization problem does not have an elegant analytical solution but can be efficiently solved using numerical methods.

However, in order to be able to fully assess the performance of the prediction algorithm an estimate of the whole posterior distribution for $ {\bf{ b } }_j^k$ might be required. A common set of numerical approaches to obtain the posterior distribution and also to perform MAP estimation is by Markov chain Monte-Carlo (MCMC) methods.

MCMC is an effective numerical approach widely used for Bayesian inference. MCMC methods were previously successfully applied to the ODM estimation problem \cite{hazelton2010statistical, park2008markov}. It is a very flexible framework that allows for studying how different components of the system influence the final prediction results. Using MCMC to solve the PFP problem we can evaluate the effects of the APC inaccuracies on passenger flow prediction. This approach also allows for an assessment of the minimum acceptable APC performance required to make useful predictions regarding passenger flow. By changing prior distribution we can also assess the effects of the historical data availability on the prediction results.

\section{Numerical results}
In order to study the performance of the proposed crowding level prediction algorithm, ARCON simulated a simple rail transit system with a single line consisting of 5 stations. Each station has a single platform with 6 boarding positions that correspond to a 6 car train. A passenger can get on the train at stations 1, 2, 3, and 4, and get off the train at stations 2, 3, 4, and 5. The station 5 is the last station of the line. No passengers must remain on the train after arriving to the station 5. 

The passengers are assumed to arrive to the platforms at constant rates according to the Poisson distribution. The arrival rates and the train headways are given in Table \ref{tb:param_sim}.

Table \ref{tb:odm_sim} contains probability values $p_{i,j}$ used in this simulation. These probabilities describe how likely it is for a passenger boarding at the $i^{\rm th}$ origin station to travel to the $j^{\rm th}$ destination station. This table is similar to an ODM, where instead of the total numbers of traveling passengers the time averaged probabilities are used. In practice these probabilities can be estimated from available historical data that describes our prior knowledge about the transit system. 

\begin{table}[!ht]
\centering
\caption{Origin-destination probabilities.}
\label{tb:odm_sim}	
	\begin{tabular}{|c|c|c|c|c|c|}
    	\hline
         \backslashbox{Origin}{Destination} & station 1 & station 2 & station 3 &  station 4 & station 5 \\
         \hline
         station 1 & & 0.2 & 0.2 & 0.3 & 0.3\\
         \hline
         station 2 & & & 0.2 & 0.4& 0.4 \\
         \hline
         station 3 & & & & 0.6 & 0.4 \\
         \hline
         station 4 & & & & & 1.0\\
         \hline
         station 5 & & & & &\\
         \hline
    \end{tabular}	
\end{table}

\begin{figure}[!ht]
\centering
\includegraphics[width=0.6\textwidth]{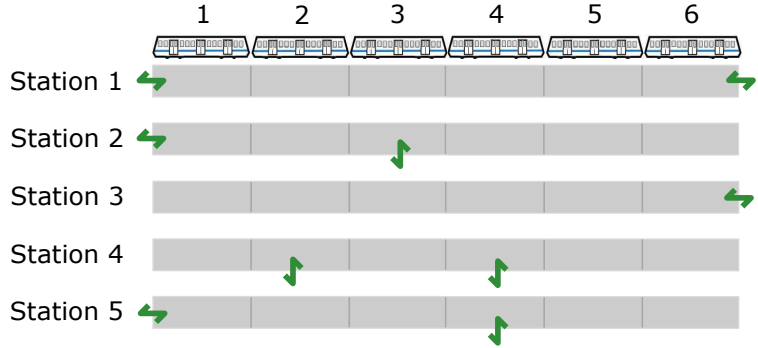}
\caption{Platform layouts used for the simulation. The access points to the platform are marked in green.}
\label{fig:station_model_sim}
\end{figure}

The layouts of the platforms at different stations are shown in Figure \ref{fig:station_model_sim}. Each platform is assumed to be divided into 6 boarding positions that correspond to each car of the train. When a passenger arrives to a platform he/she chooses one of these boarding positions and remains there until boarding the train. The passengers arrive to the platform through the access points marked in green in Figure \ref{fig:station_model_sim}. If a passenger is not a committed commuter he/she chooses the boarding position according to the distribution in (\ref{eq:boarding_dist}). For this simulation the constant $\xi$ in (\ref{eq:boarding_dist}) was taken to be equal to 1.

\begin{table}[!ht]
\centering
\caption{Passenger arrival rates and train headways.}
\label{tb:param_sim}	
	\begin{tabular}{|c|c|c|}
    	\hline
         & Passenger arrival rate& Train headway\\
         Station & [passengers / second] & [seconds]\\
         \hline
         1 & 1.5 & 180 \\
         \hline
         2 & 1.5 & 100\\
         \hline
         3 & 1.2 & 100\\
         \hline
         4 & 1.2 & 120\\
         \hline
    \end{tabular}	
\end{table}

If a passenger is a committed commuter he/she chooses the boarding position according to the platform layout at the destination. For example, if a committed passenger is traveling to the station 4, he/she will choose the boarding positions 2 or 4 with equal probability regardless of the origin station. For this simulation we assumed that the probability of a passenger being a committed commuter is $p_c = 0.3$.

The train in the simulated transit system is assumed to be equipped with an APC system. The simulated APC provides passenger counts in each train car after the train leaves a platform. Inaccuracies in the APC measurements are modeled as an additive Gaussian noise with zero mean and standard deviation $\sigma$. Zero mean noise indicates that the APC devices are assumed to provide unbiased measurements.

The transit system simulation consists of the following steps:
\begin{enumerate}
\item Generate numbers of boarding passengers for each station and each boarding position from the Poisson distribution in (\ref{eq:prior}). The parameters for this distribution are the passenger arrival rates and the train headways in Table \ref{tb:param_sim}, the travel probabilities in Table \ref{tb:odm_sim}, the platform layouts shown in Figure \ref{fig:station_model_sim}, and the chosen probability of a committed commuter $p_c = 0.3$.

\item Using Poisson distribution in (\ref{eq:prior}) as a prior come up with initial estimates of the car-level ODMs. At this moment the train has not arrived to the first station yet and no APC measurements are available. The initial crowding level predictions for the first station are generated based on the historical data only.

\item Starting at the first station for each station $j$ until the train arrives to the last station:
\begin{enumerate}
\item Simulate boarding and alighting process for each car by evaluating numbers of on-board and alighted passengers using (\ref{eq:on_board}) and (\ref{eq:alighted}).

\item By adding Gaussian random noise to the evaluated numbers of the on-board passengers synthesize APC measurements.

\item Aggregate the obtained APC measurements with the APC counts collected at the previous stations and re-estimate the elements of the car-level ODM by solving the optimization problem in (\ref{eq:map}).
\label{itm:est}

\item Estimate crowding levels for the upcoming station using (\ref{eq:crowding}).

\end{enumerate}

\item Evaluate performance of the crowding level estimation algorithm by calculating errors between the ``true" simulated crowding levels and the estimated crowding levels. 
\end{enumerate}

The model in (\ref{eq:onb-matrix}) is set up such that the algorithm uses all the available APC measurements from the first station to the current station to re-estimate the entries of the car-level ODMs. As the train proceeds forward more and more measurements become available which results in better estimation performance.

In step \ref{itm:est} of the simulation procedure we used the Metropolis-Hastings algorithm and the MCMC methods to estimate the posterior distribution in (\ref{eq:posterior}). The MAP estimates of the car-level ODMs were then found from the estimated posteriors.

Notice, that in this simulated network both the noisy APC measurements and the ``true" passenger counts are available. Thus we can evaluate the estimation performance of the proposed crowding-level prediction algorithm by comparing predictions to the truth. In practice the true passenger counts are not available, and the manual counting must be used to validate the estimation performance.

\begin{figure}[!ht]
\centering
\includegraphics[width=\textwidth]{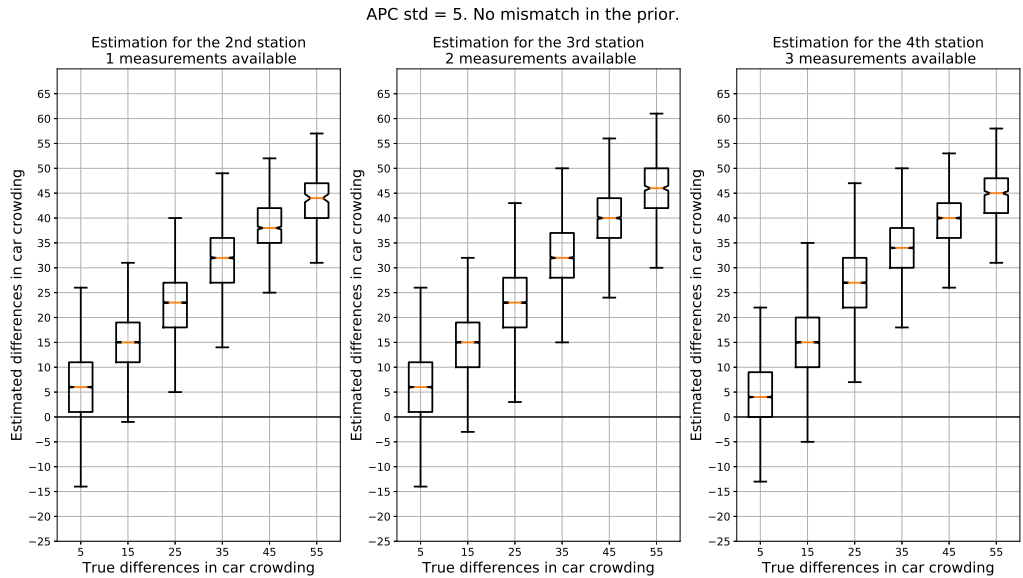}
\caption{Estimated difference in car crowding vs. true difference. The APC is unbiased with standard deviation of 5 passengers. The left panel shows crowding level prediction results before train arrives to the second station. Only one APC measurement taken after the first station is available. The center panel shows prediction results before the $3^{\rm d}$ station. Two APC measurements are available. The right panel is the results for the $4^{\rm th}$ station with three available APC measurements.}
\label{fig:mcmc_results1}
\end{figure}

\begin{figure}[!ht]
\centering
\includegraphics[width=\textwidth]{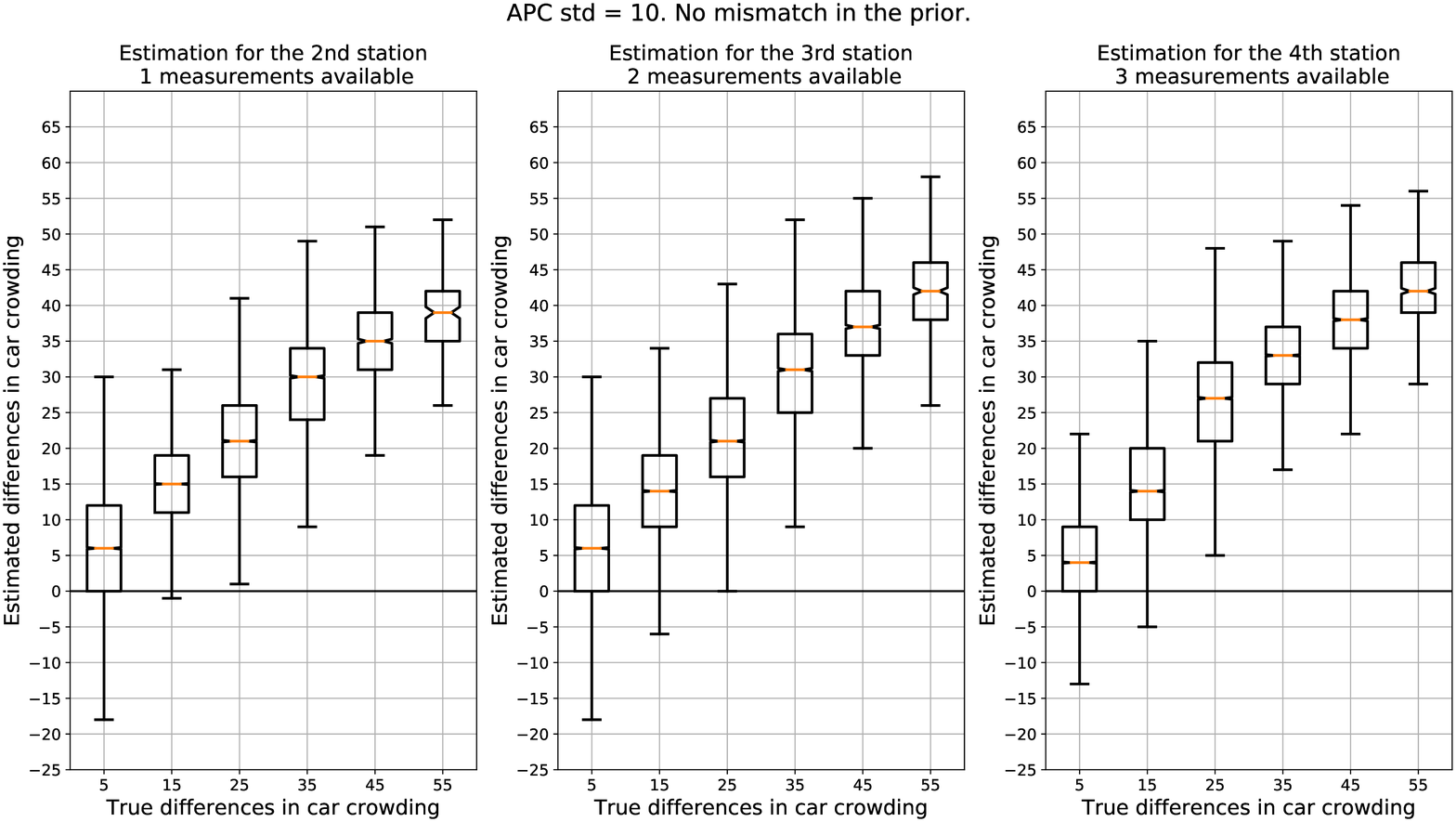}
\caption{Estimated difference in car crowding vs. true difference. The APC is unbiased with standard deviation of 10 passengers. The left panel shows crowding level prediction results before train arrives to the second station. Only one APC measurement taken after the first station is available. The center panel shows prediction results before the $3^{\rm d}$ station. Two APC measurements are available. The right panel is the results for the $4^{\rm th}$ station with three available APC measurements.}
\label{fig:mcmc_results2}
\end{figure}

\begin{figure}[!ht]
\centering
\includegraphics[width=\textwidth]{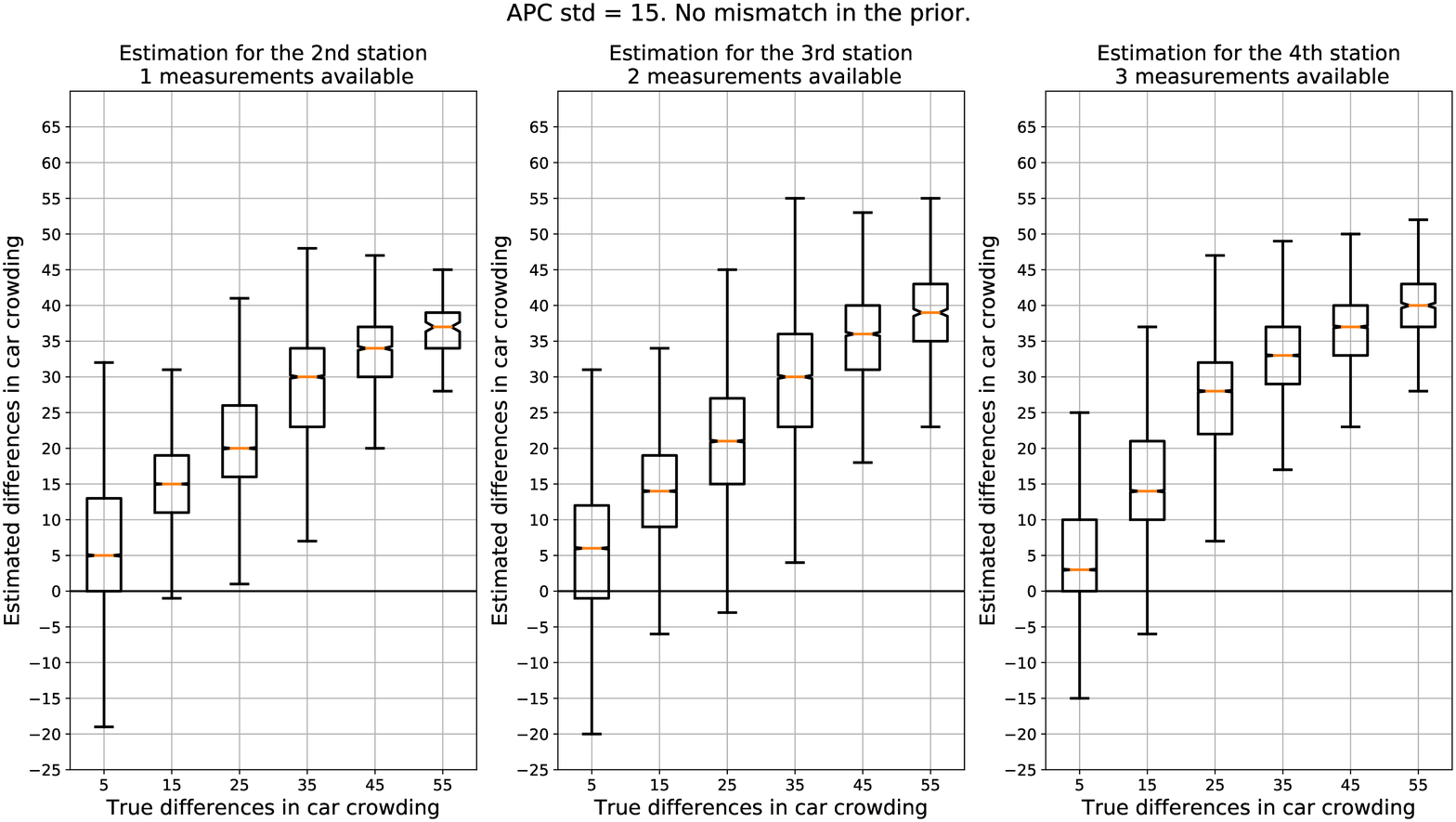}
\caption{Estimated difference in car crowding vs. true difference. The APC is unbiased with standard deviation of 15 passengers. The left panel shows crowding level prediction results before train arrives to the second station. Only one APC measurement taken after the first station is available. The center panel shows prediction results before the $3^{\rm d}$ station. Two APC measurements are available. The right panel is the results for the $4^{\rm th}$ station with three available APC measurements.}
\label{fig:mcmc_results3}
\end{figure}

\begin{figure}[!ht]
\centering
\includegraphics[width=\textwidth]{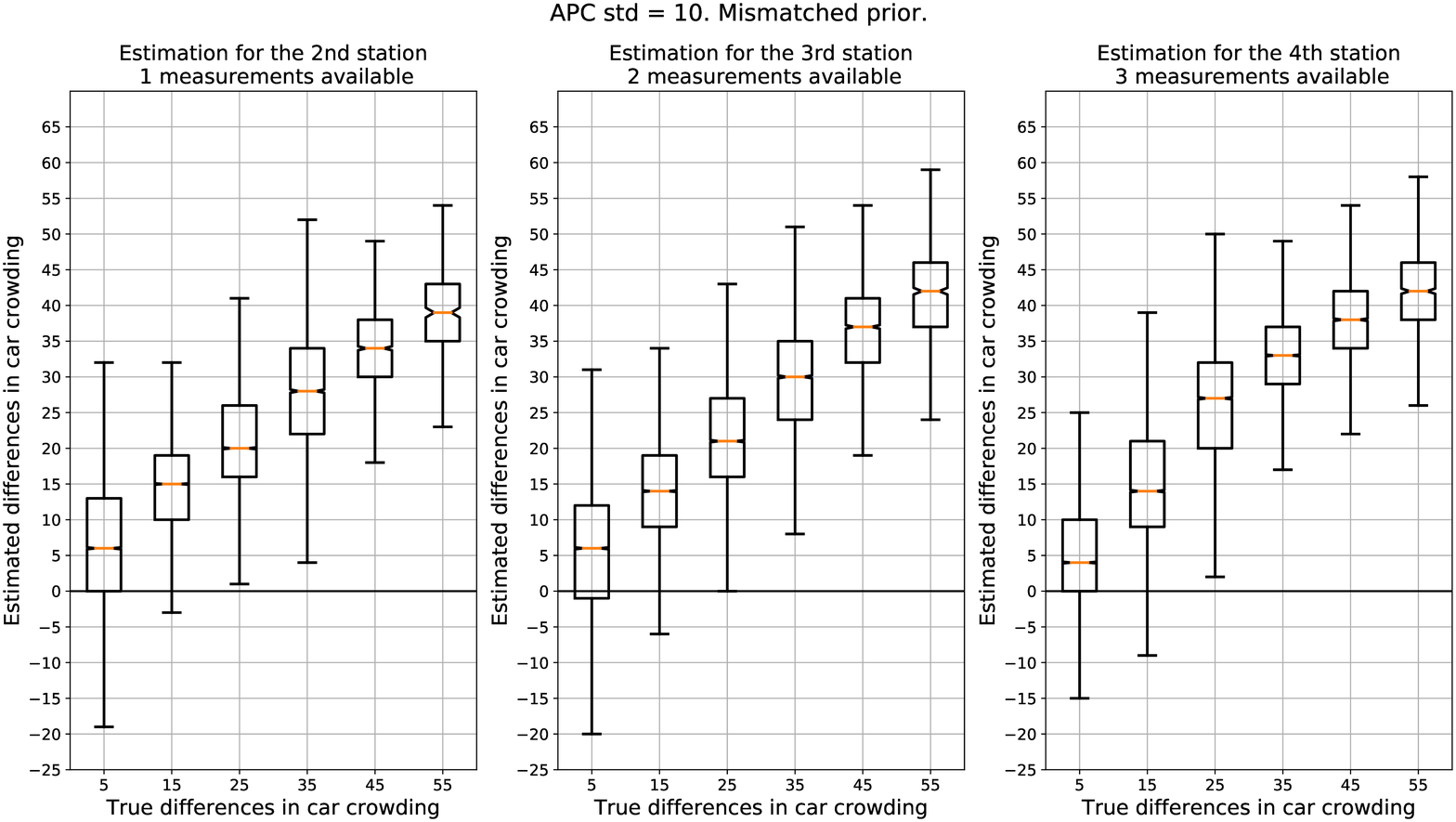}
\caption{Estimated difference in car crowding vs. true difference. The APC is unbiased with standard deviation of 10 passengers. The prior distributions for the elements of the car-level ODMs are mismatched with the true distributions that generated the passenger counts. The left panel shows crowding level prediction results before train arrives to the second station. Only one APC measurement taken after the first station is available. The center panel shows prediction results before the $3^{\rm d}$ station. Two APC measurements are available. The right panel is the results for the $4^{\rm th}$ station with three available APC measurements.}
\label{fig:mcmc_results4}
\end{figure}

\begin{figure}[!ht]
\centering
\includegraphics[width=\textwidth]{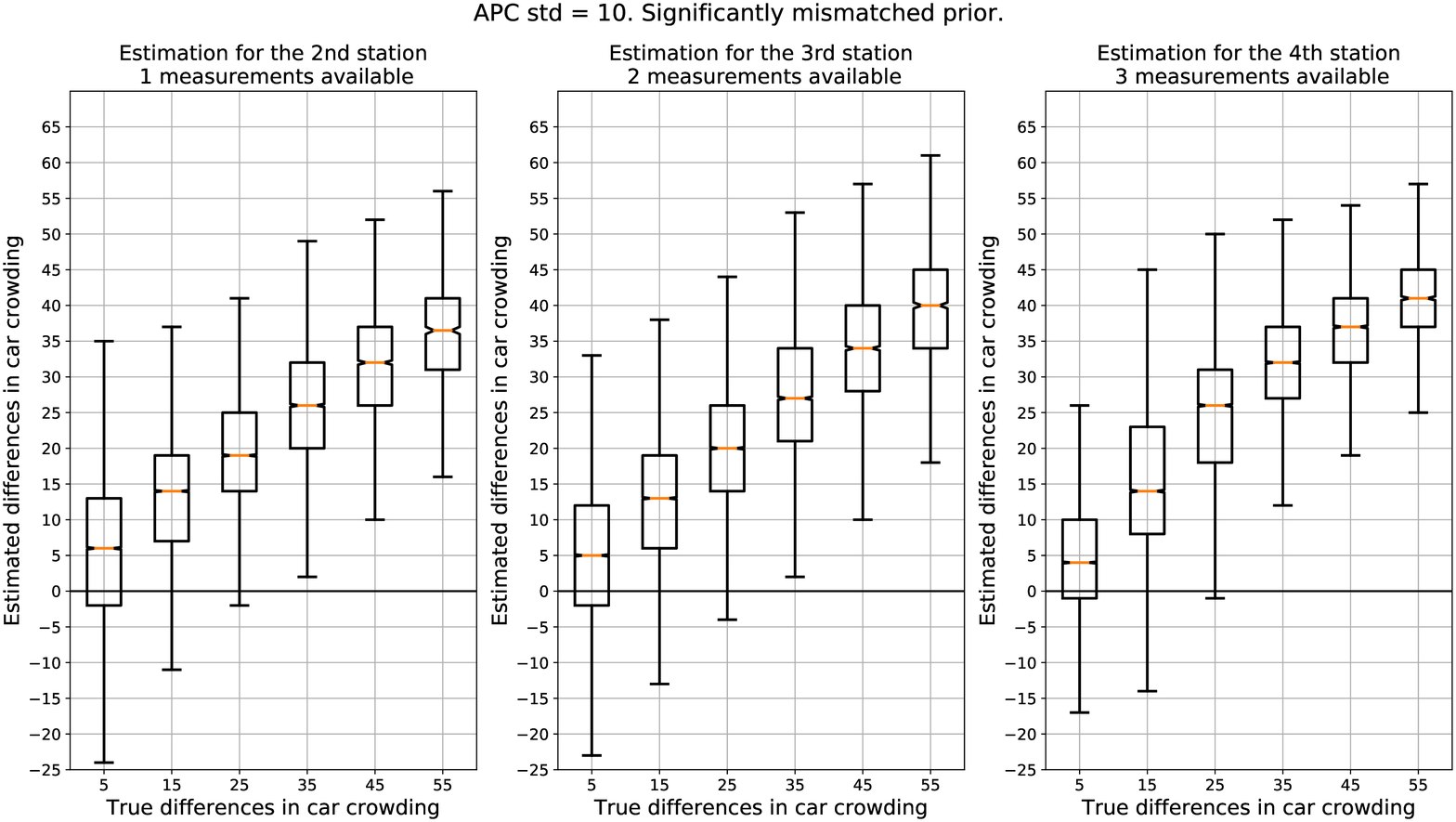}
\caption{Estimated difference in car crowding vs. true difference. The APC is unbiased with standard deviation of 10 passengers. The prior distributions for the elements of the car-level ODMs are significantly mismatched with the true distributions that generated the passenger counts. The left panel shows crowding level prediction results before train arrives to the second station. Only one APC measurement taken after the first station is available. The center panel shows prediction results before the $3^{\rm d}$ station. Two APC measurements are available. The right panel is the results for the $4^{\rm th}$ station with three available APC measurements.}
\label{fig:mcmc_results5}
\end{figure}

The estimation performance of the system is measured by examining the estimated difference in crowding levels between a pair of cars and comparing it to the true difference in crowding levels for the same pair. For example, if the true difference between the number of passengers between car 1 and car 4 is 40 and the estimated difference between the number of passengers in those cars is 52, the error is calculated as 52 - 40 = 12.  In order to collect the estimation error statistics ARCON ran 5000 Monte Carlo simulations, where at each iteration we simulated one instance of the described single line rail transit network. We present the results of these simulations as series of box-whisker plots in Figures \ref{fig:mcmc_results1}-\ref{fig:mcmc_results5}. Each figure has three panels. These panels correspond to the stations 2, 3, and 4. We do not show the prediction results for the first station because at that moment no APC measurements are available, and the quality of the crowding level prediction is determined only by the \textit{a priori} available data. The prediction results are not shown for the station 5 also because it is the last station and no passengers are assumed to board the train there.

In the presented figures the horizontal axis on each panel is the true difference between the car crowding levels at the corresponding station, while the vertical axis is the difference between the predicted car crowding levels before arriving to that station. As explained earlier these statistics are obtained by pairwise comparisons of every car to every other car in the simulated train consist. Thus the box-whisker plots show the ability of the estimation algorithm to correctly predict the difference between the crowding levels in any two cars of the train before arriving to the following station. A good prediction algorithm will have narrow boxes and whiskers along the line where the vertical axis value is equal to the horizontal axis value. 

ARCON chose such a way of presenting the performance of the designed crowding-level prediction algorithm instead of using traditional estimation error vs. truth plot, because such plot would be scenario dependent and would be difficult to interpret. The presented box-whisker plots characterize the prediction performance in general and are more suitable for analysis. Hence further results focus not on the ability to accurately predict the actual crowding level (or number of passengers)  in a particular car, but rather on reliably predicting that one car will be more crowded than the other. This is more appropriate since the disseminated output information to the passengers on the station is in the form of expected relative crowding levels between the cars, not the absolute number of passengers. 

We considered a number of scenarios with different APC qualities. For example Figure \ref{fig:mcmc_results1} shows results for an APC system with standard deviation of 5 passengers. It means that in 99\% of the time the APC counters provide an estimate that is within +/- 15 passengers from the true passenger count. The box-whisker plots in Figure \ref{fig:mcmc_results1} indicate that the algorithm can reliably detect which car will be more crowded at the following station if the true difference between the cars will be about 20-25 passengers.

By comparing the first and the third panels in Figure \ref{fig:station-model}, one can observe that as the train proceeds forward and more measurements become available, the algorithm starts providing slightly better predictions at each station. It also can be observed that the difference in crowding levels estimated by the algorithm is biased. As the true difference increases, the algorithm tends to underestimate the difference. This effect is noticeable only when the differences are large. Thus it will not affect the crowding level prediction performance, since as the difference grows the confidence in the prediction also increases despite the bias. The effect of the bias is most probably due to the nature of the Poisson prior. The bias is noticeable when the differences between two cars is large because in this case the Poisson priors with different rates are used for each car.

Figures \ref{fig:mcmc_results2} and \ref{fig:mcmc_results3} show results similar to Figure \ref{fig:mcmc_results1}, however in these scenarios APCs with larger errors were used. Figure \ref{fig:mcmc_results3} shows that when APC makes large errors, two cars have to have larger differences in crowding levels in order for the algorithm to be able to reliably predict these differences. For example, in scenario shown in Figure \ref{fig:mcmc_results3} the APC has standard deviation of 15 which can result in significantly miscounting the passengers (within +/-45 from the true number). In this case the simulation results show that if the crowding in the two cars differs by at least 30 passengers, our algorithm will be able to make a reliable prediction and tell which car will be more crowded.

In real transit network the available historical information might not accurately represent the current state of the system. Speaking in the terms of the developed statistical model, the parameters that determine the Poisson distribution in (\ref{eq:prior}), i.e. the passenger arrival rates, the travel probabilities, the probability of committed passenger etc., are approximations of the real values. Moreover, these parameters and the assumed Poisson distribution are only simplified models of the real-world phenomenas. Such models do not take into account a large number of other factors present in the real transit systems. Thus, the available prior distribution $p({\bf b}_j^k)$ used for the Bayesian estimation in (\ref{eq:map1}) might be mismatched with an actual process that produces ${\bf b}_j^k$. For example, the passenger arrival rates to the platforms are not constant but vary with time (if commuters are aware of the train schedule their arrival rate will spike closer to the train departure times). The distribution of the passengers on the platform also might take more complicated forms than the model assumed in (\ref{eq:boarding_dist}). Therefore in a real system the resulting distribution of the car-level ODM entries will not be Poisson.

To assess the performance of the proposed crowding level prediction algorithm in scenarios when the assumed prior is mismatched with an actual process we simulated two scenarios shown in Figures \ref{fig:mcmc_results4} and \ref{fig:mcmc_results5}. In these scenarios we used a distorted version of the true Poisson distribution as the prior. The distortion to the prior was introduced by randomly changing the rate parameters of the Poisson distribution in (\ref{eq:prior}). Figures \ref{fig:mcmc_results4} shows results with moderate level of the distortion, and Figure \ref{fig:mcmc_results5} shows results when a significant distortion was introduced to the prior distribution. One can observe that the distortion in the prior degrades the performance of the prediction algorithm. When a distortion is present the difference in the number of passengers between two cars must be larger in order to be reliably predicted compared to the scenarios with no distortion. These results emphasize the fact that the good prior information is important for the reliable crowding level prediction.

\section{Conclusion}
\label{ch:conclusion}
At the end of Phase I period-of-performance, ARCON has successfully completed all the proposed tasks for developing the CICD system. The two major outcomes of the Phase I efforts were (i) proof-of-concept for the accelerometer based APC device and (ii) proof-of-concept for the passenger flow prediction algorithm (PFP). The outcomes of the Phase I efforts provide the platform for a full-fledged development of APC prototype and PFP during the Phase II effort.  
 
We used theoretical model base analysis and accelerometer data from real train cars to demonstrate the feasibility of using accelerometer sensor to infer passenger loading on train cars. An accelerometer datalogger was implemented using Raspberry Pi computer and a sensor module. The accelerometer datalogger will be developed into a working APC prototype during the Phase II effort.

The prototype PFP algorithm was designed and tested on simulated train network with a single rail track. Monte Carlo experiment was used to validate the predictive performance of the algorithm. Phase I experiment demonstrated the feasibility of using the PFP algorithm for improved prediction of the estimated crowding level in train cars.
The PFP algorithm needs further refinement and modifications to accommodate a more practical multi-track train network.

\section*{Acknowledgement}
\addcontentsline{toc}{chapter}{\numberline{}Acknowledgement}%
ARCON Corporation would like to thank the Volpe National Transportation Systems Center for the SBIR grant which gave us the opportunity to work on the proposed ideas for the Crowding Information Collection and Dissemination (CICD) system. We are grateful to Dr.~Nazy Sobhi for being the technical point-of-contact for the Phase I effort. Dr.~Sobhi's guidance, feedbacks and encouragements during monthly meetings helped to steer this project in the right direction. We are also thankful to Ms.~Tammy A.\,Taylor for managing the Phase I contract logistics.

Finally, we would like to acknowledge Dr.~Laurel Paget-Seekins,  the Director of Strategic Initiatives at the Massachusetts Bay Transportation Authority (MBTA). We thank Dr.~Seekins for sharing her insights into the practical aspects of the train crowding problem. 

\bibliographystyle{ieeetr}
\bibliography{cicd} 
\end{document}